*Subject Section*

# A Method for Improving the Detection Accuracy of MSIsensor Based on Downsampling

Ji Detao[1,*], Liu Weier[2,*] and Co-Author[2]

[1]Department of XXXXXXX, Address XXXX etc., [2]Department of XXXXXXX, Address XXXX etc.

*To whom correspondence should be addressed.

## Abstract

**Motivation:** The quick brown fox jumps over the lazy dog. The quick brown fox jumps over the lazy dog. The quick brown fox jumps over the lazy dog. The quick brown fox jumps over the lazy dog. The quick brown fox jumps over the lazy dog. The quick brown fox jumps over the lazy dog.
**Results:** The quick brown fox jumps over the lazy dog. The quick brown fox jumps over the lazy dog. The quick brown fox jumps over the lazy dog. The quick brown fox jumps over the lazy dog. The quick brown fox jumps over the lazy dog. The quick brown fox jumps over the lazy dog.
**Availability:** The quick brown fox jumps over the lazy dog.
**Contact:** example@example.org
**Supplementary information:** Supplementary data are available at *Bioinformatics* online.

## 1 Introduction

Microsatelites are short repeats varying from 1-6 bp, in human DNA there are over 176 million Microsatelites locations[1]. During DNA replication, those sequence may expand or shrink due to strand slippage, often the DNA mismatch repair system can maintain genomic stability. Microsatelite instability manifests as increases or decreases in the number of repeats due to deficiency of DNA mismatch repair (MMR) system[2], and the rate of MSI in all MS locations can divided one patient sample into MSI-high or MSI-low/MSI-stable defined in the Bethesda guidelines[3].

MSI has been extensively studied in colorectal cancer, with MSI traits exhibited in 15% to 20% of colorectal cancers, and the MSI status of samples is correlated with the survival rate of stage II and stage III colorectal cancer patients [4]. Since the discovery of MSI, similar traits to those observed in colorectal cancer have also been found in gastric cancer [5] and endometrial cancer [6]. In addition to the above three types of cancer, The Cancer Genome Atlas (TCGA) project, which studied the whole genome and exome, has shown that MSI also occurs in other types of cancer[7].

The MMR mechanism can detect and identify mismatches in DNA, not only increasing the accuracy of DNA replication[8], but also reducing the occurrence of replication errors by terminating DNA during the replication process[9]. There are two key components of the MMR system that play the main role. The MutS component is responsible for detecting mismatches in DNA, while the MutL component is responsible for processing and excising the mismatches. Due to failure of MMR pathway, the mismatch replication errors persist in somatic cells and may cause cancer. In studies of human cancers[10] and mice, it has been found that the most severe MSI phenotype occurs when cells lack MSH2 or MLH1 [11], which means that MMR is unable to distinguish mutations in the absence of MutS or MutL complexes. In addition, the loss of MSH6 or PMS2 can also lead to a certain proportion of cancer development [12]. Due to the differences in the MSI phenotype caused by mutations in different MMR genes, when errors occur in the AAAG or ATAG tetranucleotide repeats at microsatellite loci, they exhibit different characteristics from the commonly recognized MSI-H phenotype. More than one-third of lung cancer, skin cancer, or bladder cancer can detect this microsatellite instability signal, which is called Elevated MS Alterations at Selected Tetranucleotide repeats (EMAST).

To detect the MSI, traditionally golden standard consist of PCR capillary electrophoresis approach and immunohistochemistry (IHC) approach[13]. During MSI-PCR sequencing, the polymerase chain reaction amplification can produce a polymerase slippage effect, which is similar to the process of MSI production and can easily lead to confusion[14];in addition, the application prospect of testing MSI status based on cfDNA is more extensive, and the use of MSI-PCR cannot obtain the MSI status information of samples with extremely low tumor content [15]. According to a study[16], when all four MMR antibodies are used for detection, the sensitivity of IHC can reach 92%, which is equivalent to MSI-PCR in sensitivity. In general, IHC can accurately identify protein truncation or degradation caused by gene mutations, but it cannot distinguish mutant proteins caused by missense mutations from wild-type polymorphisms. Most of the mutations in MSH2 result in protein truncation, so most MSH2 mutations in colorectal tumors detected by IHC show a loss of MSH2[17]. Furthermore, over one-third of MLH1 mutations involve missense mutations, which may result in a non-functional but antigenically intact mutated protein during catalysis[18]. As a result, these proteins may exhibit a false normal staining pattern when detected by IHC.

With the publication of MSIsensor, the method based on NGS data to detect MS status in samples, detection methods based on NGS data have gradually entered the vision of researchers[19]. In the early stage of MSIsensor, it was trained on 242 TCGA samples. The detection method used was to use the Chi-square test to detect the repeat frequency of bases at a specific locus in tumor and normal samples. The chi-square test is used to calculate scores for both normal and tumor samples, and according to the p-value rule, a site is determined to be an unstable site when the p-value is less than 0.05. It is worth noting that the p-value mentioned here is the first threshold used by MSIsensor. For all microsatellite loci, the Chi-square test[20] is applied to detect instability, and after correction with false discovery rate (FDR) [21], the proportion of unstable loci among all loci is defined as follows: if the proportion of unstable loci is higher than 10%, the patient sample is classified as MSI-H; if it is lower than 10% but higher than 5%, it is classified as MSI-L; if it is lower than 5%, it is classified as MSS. The 10% and 5% thresholds manually set here are the second threshold used by MSIsensor. In the following, we will mathematically prove the impact of the two thresholds on the detection results. When MSIsensor is applied to TCGA samples, most of the sequencing depths are concentrated at around 40×, which allows MSIsensor to effectively distinguish between MSI-H and MSS. However, due to the development of sequencing technology and the potential application of MSI detection in ctDNA samples, the sequencing depth of current ctDNA samples far exceeds that used when MSIsensor was developed. For example, the initial threshold of 10% used by MSIsensor in industry had to be adjusted to 20% or even 25% due to the high sequencing depth. However, this approach has significantly deviated from the original intention of using NGS data to detect MS status: the same threshold cannot be used for different samples with varying sequencing depths. Therefore, it is a perplexing issue to determine which threshold MSIsensor should use to determine the MSI status of an unknown sample.

This article will begin by introducing the principle of the Chi-square test in MSIsensor, and then use mathematical proof to demonstrate the sensitivity of the Chi-square test to frequency and the differences in Chi-square test results at different sequencing depths, in order to uncover the reason why the Chi-square test in MSIsensor is ineffective under the extremely deep sequencing depth of today. In addition, this article proposes a method based on dynamic thresholds to attempt to address the issue of poor sensitivity of MSIsensor detection under changing sequencing depths.

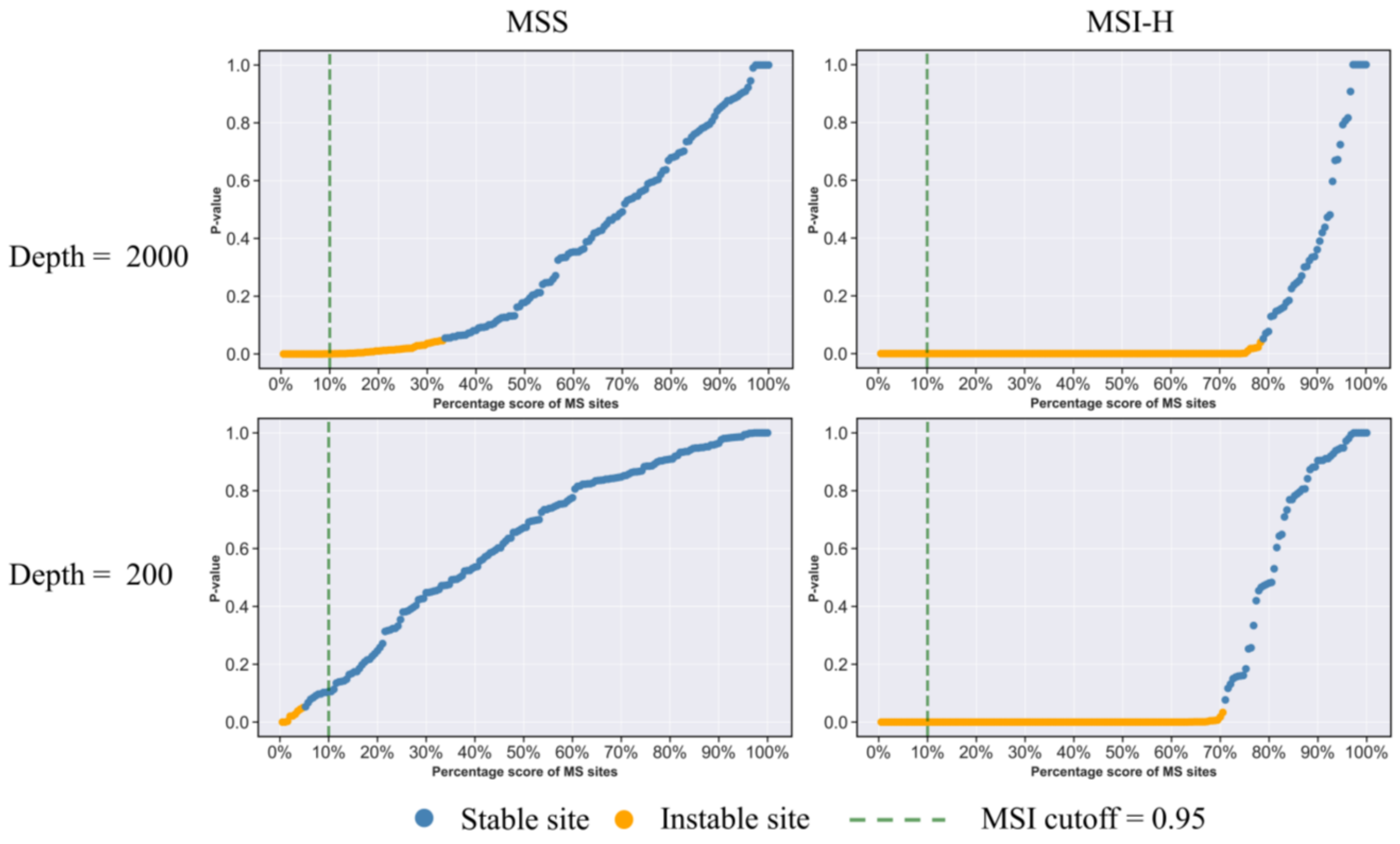

**Fig. 1. The sequencing depth is known to impact the MSI status assessment of samples using MSIsensor.** In the above figure, it can be observed that the proportions of unstable loci are different between sequencing depths of 2000x and 200x. Moreover, in MSS samples, a significant increase in instable loci in the high sequencing depth samples lead to false positive calls.

In the above figure, it can be observed that the proportions of unstable loci are different between sequencing depths of 2000x and 200x. Moreover, in MSS samples, a significant increase in instable loci in the high sequencing depth samples lead to false positive calls.

## 2 Methods

This paragraph will begin by introducing the principle of the Chi-square test in MSIsensor, and then use mathematical proof to demonstrate the sensitivity of the Chi-square test to frequency and the differences in Chi-square test results at different sequencing depths, in order to uncover the reason why the Chi-square test in MSIsensor is ineffective under the extremely deep sequencing depth of today. In addition, this article proposes a method based on dynamic thresholds to attempt to address the issue of poor sensitivity of MSIsensor detection under changing sequencing depths.

## 2.1 Chi-square test in MSIsensor

First, the scan module in Msisensor scans the reference genome (hg19, hg38, etc.) to obtain microsatellite repeat loci similar to the table below. Note that similar operations are also performed in other software such as Msisensor-pro, msings, BMSI-cast, etc. The purpose is to compress the data size for subsequent comparisons.

| Chromosome | Location | Repeat_length | Repeat_times | repeat_bases |
|---|---|---|---|---|
| 1 | 10485 | 4 | 3 | GCCC |
| 1 | 10629 | 2 | 3 | GC |
| 1 | 10652 | 2 | 3 | AG |
| 1 | 10658 | 2 | 3 | GC |
| 1 | 10681 | 2 | 3 | AF |
| … | … | … | … | … |

After running the scan module, the output.prefix_somatic file is generated, which contains information on all microsatellite loci in the reference genome. Next, the sequencing samples are compared one by one against these loci, and the repeat frequencies in the normal and tumor samples are extracted from the same loci. The following table shows the Chi-square test table for a locus in the normal and tumor samples:

| Repeat time | 1 | 2 | ... | j | ... | 100 |
|---|---|---|---|---|---|---|
| T | $T_1$ | $T_2$ | ... | $T_j$ | ... | $T_{100}$ |
| N | $N_1$ | $N_2$ | ... | $N_j$ | ... | $N_{100}$ |

In this table, assuming the repeating unit is ACG, when the repeat time is j, the T row indicates that in the tumor sample, a certain sequence segment repeats the ACG repeating sequence $T_j$ times, and the N row is similar. In MSIsensor, two hypotheses are given: $H_0$ assumes that the tumor and normal sequences belong to the same distribution, while $H_1$ assumes the opposite. Let $n = \sum_j (T_j + N_j)$, $n_T = \sum_j T_j$, $n_N = \sum_j N_j$, and $n_j = T_j + N_j$. Mathematically, this can be described in:

$$H_0 : \frac{T_j}{n_T} = \frac{N_j}{n_N}$$

$$H_1 : \frac{T_j}{n_T} \neq \frac{N_j}{n_N}$$

To determining whether two sequences belong to the same distribution, prove the following proposition true in the Chi-square test:

$$H_0' : T_j = \frac{n_j \cdot n_T}{n} \quad and \quad N_j = \frac{n_j \cdot n_N}{n}$$

When $H_0'$ is true, where $n_j = n \cdot \frac{T_j}{n_T} = n \cdot \frac{N_j}{n_N}$, it can be shown that $H_0$ is true if $\frac{T_j}{n_T} = \frac{N_j}{n_N}$.

When $H_0$ is ture, $n_j = T_j + N_j$:

$$n_j \cdot n_T = T_j \cdot n_T + N_j \cdot n_T = T_j \cdot n_T + T_j \cdot \frac{n_N}{n_T} \cdot n_T = T_j \cdot n_T + T_j \cdot n_N = T_j \cdot n$$

$$n_j \cdot n_N = T_j \cdot n_N + N_j \cdot n_N = N_j \cdot \frac{n_T}{n_N} \cdot n_N + N_j \cdot n_N = N_j \cdot n_T + N_j \cdot n_N = N_j \cdot n$$

Deviding one equation by another we can get $\frac{n_T}{n_N} = \frac{T_j}{N_j}$, which is $\frac{T_j}{n_T} = \frac{N_j}{n_N}$, means $H_0'$ is true.

Thus, it can be demonstrated that using the Chi-square test with MSI sensor to detect the MSI status of microsatellite loci is reasonable. For convenience, we will use $n_{ij}$ to represent $T_j$ or $N_j$, $n_{i\cdot}$ to represent $n_T$ or $n_N$, and $n_{\cdot j}$ to represent $n_j$. The expected value is $\text{except}(n_{ij}) = \frac{n_{\cdot j} \cdot n_{i\cdot}}{n}$, and the test statistic V is given by:

$$V = \sum_{i,j} \frac{\left(n_{ij} - \text{except}(n_{ij})\right)^2}{\text{except}(n_{ij})} = n \left( \sum_i \sum_j \frac{{n_{ij}}^2}{n_{i\cdot} \cdot n_{\cdot j}} - 1 \right)$$

When $H_0$ is true and n tends to infinity, the mathematical expectation of the test statistic V satisfies the following relation:

$$\lim_{n \to \infty} P(V \leq x) = F_k(x)$$

Here, $F_k(x)$ is the probability distribution function of the chi-square distribution:

$$F_k(x) = \frac{\gamma\left(\frac{k}{2}, \frac{x}{2}\right)}{\Gamma\left(\frac{k}{2}\right)}$$

Here, k is the degree of freedom,defined as $k = (i - 1) * (j - 1)$,where i is the number of rows and j is the number of columns. Since the degree of freedom of the data has been determined in the Chi-square test, the part of $\Gamma\left(\frac{k}{2}\right)$ is a constant, and we only need to focus on the part of $\gamma\left(\frac{k}{2}, \frac{x}{2}\right)$. When the degree of freedom k is determined, the following relationship between the expected value P and the test statistic V can be obtained:

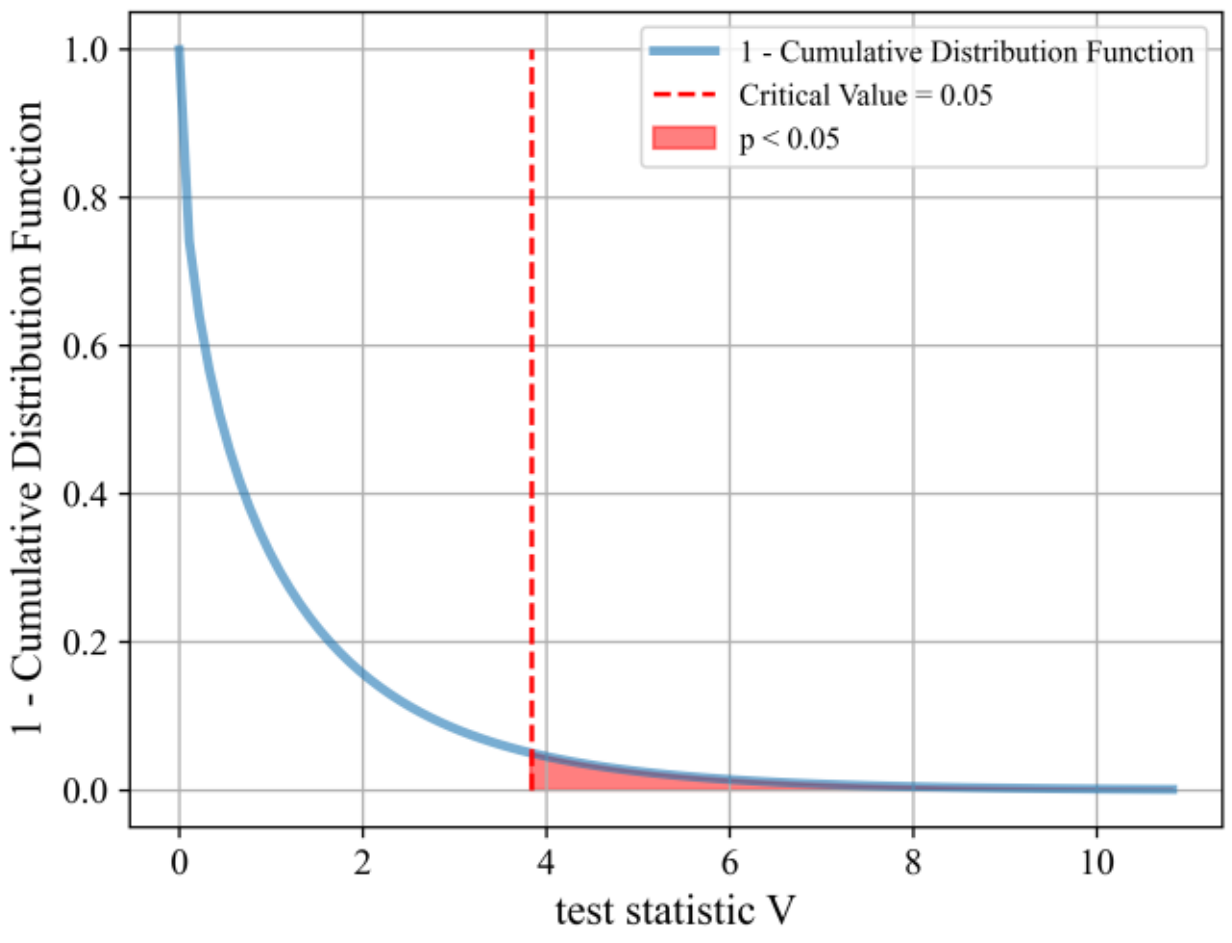

**Fig. 2. Relation between V and P.** In the above example, the degrees of freedom for the chi-square test is 1. When the calculated test statistic V is greater than 3.84, it exceeds the critical value of 0.05. In this case, the microsatellite locus is determined to be an unstable locus.

Suppose that the calculated test statistic V is 5, we can assert with a confidence level exceeding 95% that the two sequences do not conform to the same distribution. As a result, we can infer that the distributions of tumor and normal samples in MS regions, as defined by MSIsensor, are disparate.

It is noteworthy that the above conclusion is established on the basis of a sufficiently large sample size. In fact, the Chi-square test is a frequency-sensitive test, and the number of reads obtained by sequencing can directly affect the accuracy and even the threshold of the test, leading to different MS site statuses judged by the same threshold in a series of data from different platforms. Due to changes in sequencing depth, the results obtained using the same threshold and the same sample on another platform may even be completely opposite. The following section will analyze the reasons for this phenomenon and explore whether it is possible to maintain the stability of MSIsensor detection by changing some basic settings, such as normalization.

## 2.2 Instable Sequencing Depth Leads to Unstable Chi-suare Test P-Values

At the inception of MSIsensor, the sequencing depth of the TCGA dataset used was concentrated at around 30x to 50x, with an average depth of 40x. As previously demonstrated, sequencing at such stable depths could accurately distinguish MSI-H and MSS samples. However, due to the development of sequencing technologies, such as their application in MSI detection in ctDNA, sequencing depths often exceed 10,000x. Under such depths, MSIsensor's initial threshold of 10% is no longer effective. The subsequent analysis will reveal the reason for MSIsensor's insensitivity to changes in sequencing depth.

MSIsensor employs two thresholds to distinguish between MSI-H and MSS samples. The first threshold is the P-value, which can be directly calculated by the Chi-square test based on the test statistic V. In MSIsensor, the threshold for P-value is set to 0.05. The second threshold is the proportion of MSI-H sites to all MS sites across all microsatellite loci.

The following section provides evidence that P-value changes as sequencing depth changes. As previously introduced, the Chi-square test is a frequency-sensitive algorithm. Here, we demonstrate that P-value changes when the test statistic V changes, i.e., when sequencing depth increases/decreases.

Suppose now that even if we have an extremely high sequencing depth that meets the prerequisites of the Chi-square test, i.e., when $n_{i.}$, $n_{.j}$, and $n$ are all extremely large. When we simultaneously change the sequencing depth, assuming that the sequencing depth for both normal and tumor samples is increased by a factor of l, the changed sequencing depth can be expressed mathematically as $n^c_{i.}, n^c_{.j}$, and $n^c$:

$$n^c_{i.} = l \cdot n_{i.}$$
$$n^c_{.j} = l \cdot n_{.j}$$
$$n^c = l \cdot n$$

For each original sequencing data in the Chi-square table, we also performed a proportional scaling on them:

$$n^c_{ij} = l \cdot n_{ij}$$

The following relation can be easily obtained:

$$\frac{{n^c_{ij}}^2}{n^c_{i.} \cdot n^c_{.j}} = \frac{l \cdot l \cdot {n_{ij}}^2}{l \cdot n_{i.} \cdot l \cdot n_{.j}} = \frac{{n_{ij}}^2}{n_{i.} \cdot n_{.j}}$$

Assuming the statistic is $V'$, the relationship expression for $V'$ can be given as follows:

$$\begin{aligned} V' &= n^c\left(\sum_i \sum_j \frac{{n^c_{ij}}^2}{n^c_{i.} \cdot n^c_{.j}} - 1\right) \\ &= n^c\left(\sum_i \sum_j \frac{{n_{ij}}^2}{n_{i.} \cdot n_{.j}} - 1\right) \\ &= l \cdot n \cdot \left(\sum_i \sum_j \frac{{n_{ij}}^2}{n_{i.} \cdot n_{.j}} - 1\right) \\ &= l \cdot V \end{aligned}$$

When the statistical quantity V changes, as shown in Fig.2, it directly affects the size of the P-value: when l>1, that is, the sequencing depth increases by the same factor of l, the result is that the site that was unlikely to be identified as MSI-H is pushed to have an decreaseed P-value due to the increase in sequencing depth, leading to a false positive.Conversely, when l<1, i.e., the sequencing depth decreases by the same factor of l, an MSI-H site may be pushed towards the zero axis of the chart, resulting in an increase of its P-value, which leads to a false negative.

## 2.3 The Ipact of Normalization on the P-value of Ci-square Test

In fact, the issue mentioned above had already been taken into consideration during the development of MSIsensor. Since the Chi-square test is a frequency-sensitive algorithm, a straightforward and simple idea to address this issue is to flatten the frequencies. However, the following proof will demonstrate that modifying the frequencies in any way will result in changes in the P-value. Specifically, using the normalization method will lead to changes in the P-value.

First, let us introduce the solution to the aforementioned problem in MSIsensor: Normalization. The calculation of Normalization is very simple. Assuming we obtained $n_T$ and $n_N$ from the data, where $n_T \neq n_N$ (which is almost impossible to be equal), the normalization operation for each step is denoted by a prime symbol, for example, $N_j$ after Normalization becomes $N_j'$. The following operations are given:

$$\text{if } n_T > n_N\text{, then for each } N_j', N_j' = \frac{n_T}{n_N} \cdot N_j;$$
$$\text{if } n_T < n_N\text{, then for each } T_j', T_j' = \frac{n_N}{n_T} \cdot T_j;$$

After such processing, $n_N$ and $n_T$ can be normalized to be of the same magnitude, regardless of their differences. In the following proof, it will be shown that this approach still affects the P-value.

**Lemma 1:**

$$\left(T_j - \frac{n_T \cdot n_j}{n}\right)^2 = \left(N_j - \frac{n_N \cdot n_j}{n}\right)^2$$

For the Normalization step, without loss of generality, we can assume that $n_T \geq n_N$. Then we have $n_N' = n_N \cdot \frac{n_T}{n_N} = n_T$, $N_j' = \frac{n_T}{n_N} \cdot N_j$, $n_j' = T_j + N_j \cdot \frac{n_T}{n_N}$, $n' = n_N' + n_T' = 2n_T$.

**Lemma 2:**

$$\left(\frac{n_N \cdot n_j}{\mathrm{n}}\right)' = \left(\frac{n_T \cdot n_j}{\mathrm{n}}\right)'$$

**Lemma 3:**

$$\left(\frac{\left(T_j - \frac{n_T \cdot n_j}{\mathrm{n}}\right)^2}{\frac{n_T \cdot n_j}{\mathrm{n}}}\right)' = \left(\frac{\left(N_j - \frac{n_N \cdot n_j}{n}\right)^2}{\frac{n_N \cdot n_j}{\mathrm{n}}}\right)'$$

**Lemma 4:**

When $n_T \geq n_N$, we have:

$$\frac{\left(n \cdot T_j - n_T \cdot n_j\right)^2}{n \cdot n_T \cdot n_j} \leq \frac{\left(n \cdot N_j - n_N \cdot n_j\right)^2}{n \cdot n_N \cdot n_j}$$

The four lemmata above help to prove that the normalization step changes the test statistic. Next, starting from the Chi-square test statistic itself, we will demonstrate the impact of the Normalization step on the stability of its p-value.

**The p-value will increase after Normalization step:**

$$\textbf{let } \mathrm{A} = \frac{\left(n \cdot T_j - n_T \cdot n_j\right)^2}{n \cdot n_T \cdot n_j}, \mathrm{B} = \frac{\left(n \cdot N_j - n_N \cdot n_j\right)^2}{n \cdot n_N \cdot n_j}$$

If $B' \geq B$, then Lemma 3 implies that $A' = B'$ and Lemma 4 implies that $B \geq A$, thus $A' \geq A$. Therefore, $A' + B' \geq A + B$, which proves that $V'$ increases after normalization. As shown in Figure 1, the corresponding p-value(1-CDF) also increases.

Let $\xi = \frac{n_N \cdot n_j}{n}$:

$$\xi' = \frac{n_N \cdot \frac{n_T}{n_N} \cdot n_j'}{n_T + n_N \cdot \frac{n_T}{n_N}} \geq \frac{n_N \cdot \frac{n_T}{n_N} \cdot n_j'}{n_T \cdot \frac{n_T}{n_N} + n_N \cdot \frac{n_T}{n_N}} = \frac{n_N \cdot n_j'}{n} \geq \frac{n_N \cdot n_j}{n} = \xi$$

Suppose that $\alpha = \frac{n_T}{n_N} \geq 1$:

$$\begin{aligned}
\xi' &= \frac{n_N' \cdot n_j'}{n'} \\
&\leq \frac{n_N \cdot n_j'}{n} \cdot \alpha = \frac{n_N \cdot \left(T_j + N_j \cdot \frac{n_T}{n_N}\right)}{n} \cdot \alpha \\
&\leq \frac{n_N \cdot \left(T_j \cdot \frac{n_T}{n_N} + N_j \cdot \frac{n_T}{n_N}\right)}{n} \cdot \alpha \\
&= \frac{n_N \cdot n_j}{n} \cdot \alpha^2 \\
&= \xi \cdot \alpha^2
\end{aligned}$$

Then we have $\xi \leq \xi' \leq \xi \cdot \alpha^2$.

Therefore $B = \frac{(N_j - \xi)^2}{\xi}$, $B' = \frac{(N_j - \xi')^2}{\xi'} \geq \frac{\left(N_j \cdot \alpha - \frac{n_N \cdot n_j'}{n'} \cdot \alpha\right)^2}{\xi \cdot \alpha^2} = \frac{\left(N_j - \frac{n_N \cdot n_j'}{n'}\right)^2}{\xi}$.

Suppose that $c = \frac{T_j}{N_j} \geq 1$, $d = \alpha - c$;

Suppose that $a_1 = \frac{n_j'}{n_j} = \frac{T_j + \alpha \cdot N_j}{T_j + N_j} = \frac{(c+\alpha)N_j}{(c+1)N_j} = \frac{c+\alpha}{c+1} = \frac{2\alpha - d}{\alpha + 1 - d}$;

And suppose $a_2 = \frac{n'}{n} = \frac{2n_T}{n_T + n_N} = \frac{2\alpha}{\alpha+1} \geq 1$, $\beta = \frac{a_1}{a_2}$, and $\frac{\xi'}{\xi} = \frac{\frac{n_N' \cdot n_j'}{n'}}{\frac{n_N \cdot n_j}{n}} = \frac{n_N \cdot \frac{n_j'}{n'}}{n_N \cdot \frac{n_j}{n}} = \alpha \cdot \frac{n_j'}{n_j} \cdot \frac{n}{n'} = \alpha \cdot \beta$.

**When $\beta > 1$, it follows that $a_1 > a_2$:**
Hence, the following equation can be inferred from the above formula:

$$\frac{2\alpha - d}{\alpha + 1 - d} = 1 + \frac{\alpha - 1}{\alpha + 1 - d} > \frac{2\alpha}{\alpha + 1} = 1 + \frac{\alpha - 1}{\alpha + 1}$$

Due to:

$$a_1 > a_2, \ \frac{\alpha - 1}{\alpha + 1 - d} > \frac{\alpha - 1}{\alpha + 1}, \ d > 0$$

And we have:

$$d = \alpha - c > 0, \frac{n_T}{n_N} > \frac{T_j}{N_j}, \text{ we can get } n_T \cdot N_j > n_N \cdot T_j$$

At the same time, we have:

$$N_j - \frac{n_N \cdot n_j}{n} > 0 \ and \ N_j' - \frac{n_N' \cdot n_j'}{n'} > 0$$

$\because$

$$N_j' - \xi' = \alpha \cdot N_j - \alpha \cdot \beta \cdot \xi > \alpha \cdot N_j - \alpha \cdot \xi = \alpha \cdot (N_j - \xi) > 0$$

$\therefore$

$$\begin{aligned}
B' &= \frac{(N_j' - \xi')^2}{\xi'} \\
&= \frac{(\alpha \cdot N_j - \alpha \cdot \beta \cdot \xi)^2}{\alpha \cdot \beta \cdot \xi} \\
&> \frac{(\alpha \cdot n_{Nj} - \alpha \cdot \xi)^2}{\alpha \cdot \beta \cdot \xi} = \frac{\alpha}{\beta} \cdot B
\end{aligned}$$

$\because$ $$\xi' = \alpha \cdot \beta \cdot \xi \leq \xi \cdot \alpha^2$$
$\therefore$ $$\frac{\alpha}{\beta} \geq 1$$
$$B' > B$$

**When $\beta = 1$, it follows that $a_1 = a_2$, At this point $d = 0$ or $\alpha = 1$:**

When $d = 0$, $\alpha = c$, $N_j - \frac{n_N \cdot n_j}{n} = 0$, $N_j' - \frac{n_N' \cdot n_j'}{n'} = 0$, and we have $B' = B = 0$.

When $\alpha = 1$, $n_N = n_T$, the values of each item will not change during the execution of the normalization step, $B' = B$.

**When $\beta < 1$, it follows that $a_1 < a_2$:**
At this time, we have:

$$d < 0, \alpha < c, \frac{n_T}{n_N} < \frac{T_j}{N_j}$$

It can be inferred that:

$$n_T \cdot N_j < n_N \cdot T_j$$

And we have:

$$N_j - \frac{n_N \cdot n_j}{n} < 0, N_j' - \frac{n_N' \cdot n_j'}{n'} < 0$$

$\because$

$$\begin{aligned}
0 > N_j' - \xi' &= \alpha \cdot n_{Nj} - \alpha \cdot \beta \cdot \xi \\
&> \alpha \cdot \beta \cdot n_{Nj} - \alpha \cdot \beta \cdot \xi
\end{aligned}$$

$\therefore$

$$\begin{aligned}
B' &= \frac{(N_j' - \xi')^2}{\xi'} \\
&= \frac{(\alpha \cdot N_j - \alpha \cdot \beta \cdot \xi)^2}{\alpha \cdot \beta \cdot \xi} \\
&> \frac{(\alpha \cdot \beta \cdot N_j - \alpha \cdot \beta \cdot \xi)^2}{\alpha \cdot \beta \cdot \xi} \\
&= \alpha \cdot \beta \cdot B
\end{aligned}$$

$\therefore$ $$B' > B$$

In summary, if $B' \geq B$, then it follows that $A' + B' \geq A + B$.
Moreover, due to:

$$V = \sum_j \frac{(n \cdot T_j - n_T \cdot n_j)^2}{n \cdot n_T \cdot n_j} + \frac{(n \cdot N_j - n_N \cdot n_j)^2}{n \cdot n_N \cdot n_j} = \sum_j (A_j + B_j)$$

Thus, it can be demonstrated that after performing normalization, the V value increases, and correspondingly, the P-value decrease.

## 3 Results

The quick brown fox jumps over the lazy dog. The quick brown fox jumps over the lazy dog. The quick brown fox jumps over the lazy dog. The quick brown fox jumps over the lazy dog. The quick brown fox jumps over the lazy dog. The quick brown fox jumps over the lazy dog. The quick brown fox jumps over the lazy dog. The quick brown fox jumps over the lazy dog. The quick brown fox jumps over the lazy dog. The quick brown fox jumps over the lazy dog. The quick brown fox jumps over the lazy dog. The quick brown fox jumps over the lazy dog. The quick brown fox jumps over the lazy dog. The quick brown fox jumps over the lazy dog. The quick brown fox jumps over the lazy dog. The quick brown fox jumps over the lazy dog. The quick brown fox jumps over the lazy dog. The quick brown fox jumps over the lazy dog. The quick brown fox jumps over the lazy dog. The quick brown fox jumps over the lazy dog. The quick brown fox jumps over the lazy dog. The quick brown fox jumps over the lazy dog. The quick brown fox jumps over the lazy dog. The quick brown fox jumps over the lazy dog. The quick brown fox jumps over the lazy dog.

## 2.4 Data Structure This is Heading 2 style this is heading 2 style

The quick brown fox jumps over the lazy dog. The quick brown fox jumps over the lazy dog. The quick brown fox jumps over the lazy dog. The quick brown fox jumps over the lazy dog. The quick brown fox jumps over the lazy dog. The quick brown fox jumps over the lazy dog. The quick brown fox jumps over the lazy dog. The quick brown fox jumps over the lazy dog. The quick brown fox jumps over the lazy dog. The quick brown fox jumps over the lazy dog. The quick brown fox jumps over the lazy dog. The quick brown fox jumps over the lazy dog. The quick brown fox jumps over the lazy dog.

### 3.1.1 This is heading 3 style

The quick brown fox jumps over the lazy dog. The quick brown fox jumps over the lazy dog. The quick brown fox jumps over the lazy dog. The quick brown fox jumps over the lazy dog.

(1) The quick brown fox jumps over the lazy dog. The quick brown fox jumps over the lazy dog.

(2) The quick brown fox jumps over the lazy dog. The quick brown fox jumps over the lazy dog.

(3) The quick brown fox jumps over the lazy dog.

(4) The quick brown fox jumps over the lazy dog. The quick brown fox jumps over the lazy dog.

(5) The quick brown fox jumps over the lazy dog. The quick brown fox jumps over the lazy dog.

(6) The quick brown fox jumps over the lazy dog. The quick brown fox jumps over the lazy dog.

The quick brown fox jumps over the lazy dog. The quick brown fox jumps over the lazy dog. The quick brown fox jumps over the lazy dog. The quick brown fox jumps over the lazy dog. The quick brown fox jumps over the lazy dog. The quick brown fox jumps over the lazy dog. The quick brown fox jumps over the lazy dog. The quick brown fox jumps over the lazy dog. The quick brown fox jumps over the lazy dog. The quick brown fox jumps over the lazy dog. The quick brown fox jumps over the lazy dog. The quick brown fox jumps over the lazy dog. The quick brown fox jumps over the lazy dog. The quick brown fox jumps over the lazy dog. The quick brown fox jumps over the lazy dog. The quick brown fox jumps over the lazy dog. The quick brown fox jumps over the lazy dog. The quick brown fox jumps over the lazy dog.

- The quick brown fox jumps over the lazy dog. The quick brown fox jumps over the lazy dog.
- The quick brown fox jumps over the lazy dog. The quick brown fox jumps over the lazy dog.
- The quick brown fox jumps over the lazy dog. The quick brown fox jumps over the lazy dog.

- The quick brown fox jumps over the lazy dog. The quick brown fox jumps over the lazy dog.
- The quick brown fox jumps over the lazy dog. The quick brown fox jumps over the lazy dog.

The quick brown fox jumps over the lazy dog. The quick brown fox jumps over the lazy dog. The quick brown fox jumps over the lazy dog.

## 2.5 Unnumbered list style

The quick brown fox jumps over the lazy dog. The quick brown fox jumps over the lazy dog. The quick brown fox jumps over the lazy dog.

The quick brown fox jumps over the lazy dog. The quick brown fox jumps over the lazy dog.

The quick brown fox jumps over the lazy dog. The quick brown fox jumps over the lazy dog.

The quick brown fox jumps over the lazy dog. The quick brown fox jumps over the lazy dog.

The quick brown fox jumps over the lazy dog. The quick brown fox jumps over the lazy dog.

The quick brown fox jumps over the lazy dog. The quick brown fox jumps over the lazy dog.

$$Pr(\mu) = a_\mu \Big/ \sum_j a_j \qquad (1)$$

The quick brown fox jumps over the lazy dog. The quick brown fox jumps over the lazy dog. The quick brown fox jumps over the lazy dog. The quick brown fox jumps over the lazy dog. The quick brown fox jumps over the lazy dog. The quick brown fox jumps over the lazy dog.

**Fig. 1. Relation between τ and *t*.** This example has only two continuous Steppers, $S_1$ and $S_2$.

The quick brown fox jumps over the lazy dog. The quick brown fox jumps over the lazy dog. The quick brown fox jumps over the lazy dog. The quick brown fox jumps over the lazy dog. The quick brown fox jumps over the lazy dog. The quick brown fox jumps over the lazy dog. The quick brown fox jumps over the lazy dog. The quick brown fox jumps over the lazy dog. The quick brown fox jumps over the lazy dog. The quick brown fox jumps over the lazy dog. The quick brown fox jumps over the lazy dog. The quick brown fox jumps over the lazy dog. The quick brown fox jumps over the lazy dog. The quick brown fox jumps over the lazy dog.

The quick brown fox jumps over the lazy dog. The quick brown fox jumps over the lazy dog. The quick brown fox jumps over the lazy dog. The quick brown fox jumps over the lazy dog. The quick brown fox jumps over the lazy dog. The quick brown fox jumps over the lazy dog. The quick brown fox jumps over the lazy dog. The quick brown fox jumps over the lazy dog. The quick brown fox jumps over the lazy dog. The quick brown fox jumps over the lazy dog. The quick brown fox jumps over the lazy dog. The quick brown fox jumps over the lazy dog. The quick brown fox jumps over the lazy dog. The quick brown fox jumps over the lazy dog. The quick brown fox jumps over the lazy dog.

The quick brown fox jumps over the lazy dog. The quick brown fox jumps over the lazy dog. The quick brown fox jumps over the lazy dog. The quick brown fox jumps over the lazy dog. The quick brown fox jumps over the lazy dog. The quick brown fox jumps over the lazy dog. The quick brown fox jumps over the lazy dog. The quick brown fox jumps over the lazy dog. The quick brown fox jumps over the lazy dog. The quick brown fox jumps over the lazy dog. The quick brown fox jumps over the lazy dog. The quick brown fox jumps over the lazy dog. The quick brown fox jumps over the lazy dog. The quick brown fox jumps over the lazy dog. The quick brown fox jumps over the lazy dog. The quick brown fox jumps over the lazy dog. The quick brown fox jumps over the lazy dog. The quick brown fox jumps over the lazy dog. The quick brown fox jumps over the lazy dog. The quick brown fox jumps over the lazy dog. The quick brown fox jumps over the lazy dog. The quick brown fox jumps over the lazy dog. The quick brown fox jumps over the lazy dog. The quick brown fox jumps over the lazy dog. The quick brown fox jumps over the lazy dog.

**Table 1.** Benchmark results of the cascade oscillators model

| \|S\| | Predicted cost | Timing | Predicted speed | Speed |
|---|---|---|---|---|
| 1 | S219.20(100%) | 68m43s | 1.00 | 1.00 |
| 2 | $2^9$.10+$2^{19}$.10(~50%) | 35m13s | 2.00 | 1.95 |
| 4 | $2^{19}$.20(100%) | 68m43s | 1.00 | 1.00 |
| 10 | $2^9$.10+$2^{19}$.10(~50%) | 35m13s | 2.00 | 1.95 |
| 20 | $2^{19}$.20(100%) | 68m43s | 1.00 | 9.5 |

This is table foot note sample text This is table foot note sample text This is table foot note sample text

The quick brown fox jumps over the lazy dog. The quick brown fox jumps over the lazy dog. The quick brown fox jumps over the lazy dog. The quick brown fox jumps over the lazy dog. The quick brown fox jumps over the lazy dog. The quick brown fox jumps over the lazy dog. The quick brown fox jumps over the lazy dog. The quick brown fox jumps over the lazy dog.The quick brown fox jumps over the lazy dog. The quick brown fox jumps over the lazy dog. The quick brown fox jumps over the lazy dog. The quick brown fox jumps over the lazy dog. The quick brown fox jumps over the lazy dog. The quick brown fox jumps over the lazy dog.The quick brown fox jumps over the lazy dog.

## Acknowledgements

The quick brown fox jumps over the lazy dog. The quick brown fox jumps over the lazy dog. The quick brown fox jumps over the lazy dog. The quick brown fox jumps over the lazy dog. The quick brown fox jumps over the lazy dog. The quick brown fox jumps over the lazy dog.

## Funding

This work has been supported by the .....

*Conflict of Interest:* none declared.